\documentclass[prl,twocolumn,amsmath,amssymb,groupedaddress,superscriptaddress]{revtex4}
\usepackage{graphics,color}
\usepackage{amsmath} 
\usepackage{amssymb,mathtools}
\usepackage[dvips]{graphicx}
\usepackage{float,graphicx}
\usepackage{subfigure,hyperref}
\usepackage{epstopdf}
\usepackage{epstopdf}
\usepackage{soul} 
\usepackage{epstopdf}
\newcommand{\beq}{\begin{equation}}
\newcommand{\eeq}{\end{equation}}
\newcommand{\beqa}{\begin{eqnarray}}

\newcommand{\eeqa}{\end{eqnarray}}

\def\be{\begin{equation}}

\def\ee{\end{equation}}

\def\ttr{\textmd{tr}}

\def\eeq{\textmd{eq}}
\def\sst{\scriptsize\textmd{st}}

\def\tth{\textmd{th}}

\def\Rre{\textmd{Re}}
\def\Iim{\textmd{Im}}

\def\be{\begin{equation}}
\def\ee{\end{equation}}
\def\bea{\begin{eqnarray}}
\def\eea{\end{eqnarray}}

\usepackage{bm}
\usepackage{graphicx}

\begin{document}
\title{Open quantum systems in thermal non-ergodic environments}
\author{Carlos A. Parra-Murillo}
\author{Max Bramberger}
\affiliation{Department of Physics and Arnold Sommerfeld Center for Theoretical
Physics, Ludwig-Maximilians-University Munich, Germany}
\author{Claudius Hubig}
\affiliation{Max-Planck-Institut f\"ur Quantenoptik, Hans-Kopfermannstr.~1, 85748, Garching, Germany}
\author{In\'es De Vega}
\affiliation{Department of Physics and Arnold Sommerfeld Center for Theoretical
Physics, Ludwig-Maximilians-University Munich, Germany}

\begin{abstract}
The dynamics of an open system crucially depends on the correlation function of its environment, $C_B(t)$. We show that for thermal non-Harmonic environments $C_B(t)$ may not decay to zero but to an offset, $C_0>0$. The presence of such offset is determined by the environment eigenstate structure, and whether it fulfills or not the eigenstate thermalization hypothesis. Moreover, we show that a $C_0>0$ could render the weak coupling approximation inaccurate and prevent the open system to thermalize. Finally, for a realistic environment of dye molecules, we show the emergence of the offset by using matrix product states (MPS), and discuss its link to a 1/f noise spectrum that, in contrast to previous models, extends to zero frequencies. Thus, our results may be relevant in describing dissipation in quantum technological devices like superconducting qubits, which are known to be affected by such noise. 
\end{abstract}

\maketitle

The performance of quantum computers and other quantum technological devices critically relies on our ability to preserve their quantum mechanical character despite the presence of different types of noise. In general the  system of interest ($H_S$) is coupled to the noise source or environment ($H_E$)  through an interaction of the form $H_I=BS$, where $B$ and $S$ are the environment and system coupling operators, respectively \cite{breuerbook,devega2015c,rivas2011a,rivas2014,breuer2015,li2018}. The dynamics of the system mean values, described with its reduced density operator $\rho_S(t)=\ttr_E\{\rho(t)\}$ ($\rho(t)$ being the total state at the time $t$), is strongly conditioned by the different moments of the noise operator $B$ with respect to the environment initial state $\rho_E$. When the environment is a set of harmonic oscillators in a Gaussian state and  linearly coupled to the system, its statistics is Gaussian and all higher order moments are either zero or can be written in terms of the second order one, the correlation function
\bea
C_B(t)=\ttr_E\{\tilde{B}(t)\tilde{B}(0)\rho_E\},
\label{second}
\eea
via the Wick's theorem. Here we have defined a renormalized operator $\tilde B=B(t)-\ttr\{B(0)\rho_B\}$, and renormalized the system Hamiltonian accordingly $\tilde{H}_S=H_S+S\ttr\{B(0)\rho_E\}$. We assume in addition that the environment is initially in a thermal state $\rho_E=\rho_E^\tth=e^{-\beta H_E}/Z_E$, with  $Z_E$ the partition function and $\beta$ the inverse temperature. Thus, $[\rho_E,H_E]=0$ and therefore  $C_B(t_1,t_2)=\ttr_E\{\tilde{B}(t_1)\tilde{B}(t_2)\rho_E(0)\}$ depends only on the time difference $t=t_1-t_2$, i.e.  $C_B(t_1,t_2)\equiv C_B(t)$.  
Recent progress in noise spectroscopy in superconducting qubits \cite{norris2016,sung2019} has revealed the non-Gaussian nature of their environment, in particular the $1/\textmd{f}$ noise produced by surface impurities  \cite{seidler1996,faoro2004}. The non-Gaussianity affects the reduced dynamics of the system in the strong coupling regime, when higher order moments are relevant. Yet, in the weak coupling (when $g\approx ||H_I||$ is small compared to all other energy scales) $\rho_S(t)$ is dominated by the correlation function (\ref{second}) which suggests that Gaussian and non-Gaussian statistics may be hard to distinguish. 

Here, we argue that this might not be the case. In detail, we show that: (i) non-Gaussian statistics may reveal a distinguising feature already in the weak coupling regime, giving rise to a non-decaying correlation, such that $\lim_{t\rightarrow\infty}C_B(t)=C_0$. The emergence of this offset $C_0$ is linked to the statistical properties of the environment eigenstates; and (ii) the offset dramatically affects the accuracy of the weak coupling approximation, as well as the derivation of the related Lindblad equation.
There are already some evidences in this direction: In classical statistical physics 
it is known that for a system of non-interacting or weakly interacting particles the long time decay of the correlation is linked to ergodicity \cite{khinchin1949,mazur1963,marconi2008}. In other words, $\lim_{T\rightarrow\infty}\int_0^TB(t)dt=\textmd{av}[B]$, iff,  $\lim_{t\rightarrow\infty}\textmd{av}[B(t)B(0)]=\textmd{av}[B^2]$, where $\textmd{av}[\cdots]$ denotes an ensemble average. 
For a system in equilibrium, $\textmd{av}[B(t)]=\textmd{av}[B]$ and therefore the above condition can be written as $\lim_{t\rightarrow\infty}\textmd{av}[\tilde{B}(t)\tilde{B}]=0$. Furthermore, a \textit{sufficient condition} for the existence of a well-defined weak coupling limit is that there is an $\epsilon>0$, such that \cite{davies1976,davies1974}
\bea
\int_0^\infty dt|C_B(t)|(1+|t|)^\epsilon< \infty.
\label{davies}
\eea
Since $\textmd{av}[\tilde{B}(t)\tilde{B}]$ is the classical $C_B$, when the ergodicity condition is not fulfilled the condition in Eq. (\ref{davies}) is not fulfilled either and the convergence of a weak coupling expansion is not ensured. Here we discuss the conditions for the emergence of a non-decaying correlation function in a quantum system, and show how this is related to a failure of the standard weak coupling limit when describing an impurity coupled to it. Moreover, we show that the decay of $C_B(t)$ is linked in some cases to the fulfilment of the eigenstate thermalization hypothesis (ETH). Finally, we illustrate the consequences of a hybrid environment formed by dye molecules \cite{montina2008,huang2010,zaman2014}, which we relate to the extension of the $1/\textmd{f}$ spectrum to zero frequencies.
%

\textit{Environment correlation function--}
General statements on the environment correlation function can be established when representing it in its eigenbasis. Let $\rho_E=\sum_{kk}\eta_k|\epsilon_k\rangle\langle\epsilon_k|$ be an equilibrium state of the environment, with $|\epsilon_k\rangle$ its eigenstates. Thus, equation~(\ref{second}) can be written as
\bea
C_B(t)=\sum_{l,k,\omega_{kl}\neq 0}\eta_k|{B}^{kl}|^2e^{i\omega_{kl}t}+C_0,
\label{eq:2}
\eea
where $\omega_{kl}=\epsilon_k-\epsilon_l$, ${B}^{kl}=\langle \epsilon_l|B|\epsilon_k\rangle$, and 
\bea
C_0= \sum_{k}\eta_k(B^{kk})^2-\big(\sum_{k}\eta_k{B}^{kk}\big)^2 + d_0\,.
\label{offset}
\eea
It is expected that, for a sufficiently large environment, the time-dependent term in Eq.~(\ref{eq:2}) decays for $t\gg 1/{\rm min}\{|\omega_{kl}|\}$. The correlation function, however, may contain a finite offset $C_0> 0$, which is non-vanishing if: (a) at least two (or more) eigenstates of $H_E$ are degenerated (a contribution casted in the term $d_0$). Unless the system exhibits accidental degeneracy \cite{hou2017,sakurai}, degeneracies are rare as they are easily removed by the presence of a very small perturbation.
(b) Alternatively, we require that $\sum_{k}\eta_k({B}^{kk})^2>\langle B\rangle^2$ (assuming $d_0=0$), for which a necessary condition is that the sum in $k$ includes at least two eigenstates for which $\eta_k ({B}^{kk})^2\neq 0$ is fulfilled.

\textit{Statistical typicality--}
An environment obeying statistical typicality \cite{gaspard1999b,berry1977,shapiro1988} violates condition (b). This relies on the fact that for all $k$, $B^{kk}\approx\ttr_E\{\rho^\tth_EB\}$, leading to a zero offset due to the normalization condition $\sum\eta_k=1$. 
This conjecture can be justified if the bath is classically chaotic, and therefore is linked to the ergodic condition described in the introduction \cite{gaspard1999b}. 

\textit{Eigenstate thermalization hypothesis (ETH)--}
The correlation function will also decay to zero under ETH conditions. 
To show this, let us first 
rewrite Eq.~(\ref{eq:2}) in terms of $\omega_{kl}$ and $E_{kl}=\epsilon_k+\epsilon_l$ as $C_B(t)=\sum_{E_{kl}}\eta(E_{kl})C_{E_{kl}}(t)$, with 
\bea
C_{E_{kl}}(t)=\sum_{\omega_{kl}}\eta(\omega_{kl})|\tilde{B}(\omega_{kl},E_{kl})|^2e^{i\omega_{kl}t}\,.
\label{CE}
\eea 
Note that here we have considered the renormalized operator $\tilde{B}$ and assumed no degeneracies, so that $B^{kl}$ is fully determined by the eigenenergies $\epsilon_k$ and $\epsilon_l$. Moreover,  the coefficients of the thermal state have been rewritten as $\eta_k=\eta(E_{kl})\eta(\omega_{kl})/Z_E$, with $\eta(E_{kl})=e^{-\beta E_{kl}/2}$ and $\eta(\omega_{kl})=e^{-\beta \omega_{kl}/2}$. In the continuum limit of Eq. (\ref{CE}), $C_E(t)$ can be considered as the Fourier transform of $F_{E}(\omega)=\eta(\omega)|\tilde{B}(\omega,E)|^2$. Thus, following the Paley-Wiener theorem it is possible to state that if $F_{E}(\omega)$ is smooth and $N$ times differentiable in a finite support $\omega\in[-\omega_m,\omega_m]$, with $2\omega_m$ a finite bandwidth, then there is always a coefficient $C_N$ for which 
\bea
C_{E}(t)\le C_N (1+|t|)^{-N},
\label{decayC}
\eea
i.e. the correlation function decays to zero. We note that since $\eta(\omega)$ is smooth function, the smoothness condition for $F_{E}(\omega)$ is determined by $|\tilde{B}(\omega,E)|^2$. 
Following the ETH ansatz, the matrix element of the observable $\tilde{B}$ between two eigenstates can be written as $\tilde{B}^{kl}={\mathcal B}(\frac{1}{2}E_{kl})\delta_{kl}+e^{-S(\frac{1}{2}E_{kl})/2}f_0(E_{kl},\omega_{kl})R_{kl}$  \cite{dalessio2016,srednicki1999} where $S(E)$ is the thermodynamic entropy at energy $E$, and $R_{kl}$ is a random real or complex variable with zero mean and unit variance ($\overline{R^2_{kl}}=1$ or $\overline{|R_{kl}|^2}=1$, with $\overline{\cdots}$ representing the noise average). The function ${\mathcal B}(E)$ is the expectation value of the observable in the microcanonical ensemble at energy $E$, which in the thermodynamic limit corresponds its canonical average and therefore is a constant.
Considering in $C_B(t)=\sum_{l,k}\eta_k|\tilde{B}^{kl}|^2e^{i\omega_{kl}t}$ and performing the noise average, we find that it can be written as a function of Eq.~(\ref{CE}) but now with $|\tilde{B}(\omega_{kl},E_{kl})|^2=f^2_0(\omega_{kl},E_{kl})e^{-S(\frac{1}{2}E_{kl})}$. According to the ETH ansatz, $f_0$ is a smooth function, and therefore $C_B(t)$ will decay to zero according to Eq.~(\ref{decayC}). 

Hence, systems whose classical analogue is chaotic or more generally, for which ETH is suitable will present no offset. However, the opposite is not necessarily true: the fact that a system is integrable does not mean that the autocorrelation function of some of its observables may have an offset. An example of this is an environment of harmonic oscillators where $B=\sum_\lambda g_\lambda (b_\lambda\pm b_\lambda^\dagger)$, with $b_\lambda(b_\lambda^\dagger)$ annihilation (creation) operators \cite{feynman1963a}. If $\rho_E$ is in equilibrium, it is fulfilled that $B^{kk}=0$ and therefore there is no offset (see Supplementary Material (SM) for details).

\textit{Weak coupling approximation--} 
The ETH ansatz leads to a correlation given by Eq. (\ref{decayC}) that fulfils the convergence criteria (\ref{davies}), with an $\epsilon$ such that $N>\epsilon$ (notice that $C_B(t)$ is just a linear combination of $C_E(t)$).
A correlation function with an offset will not only dissatisfy Eq.~(\ref{davies}), but it will also lead to an ill-defined weak coupling master equation and to the absence of a Lindblad limit. To see this, we analyse the weak coupling master equation 
up to second order in the coupling parameter: $\partial_t\rho_s (t)=\int_0^t d\tau C_B(t-\tau) [ S(\tau) \rho_s (t-\tau),S(t)]+{\textmd h.c.}$, where $C_B(t)$ is given by Eq. (\ref{eq:2}). We now consider the spectral decomposition $S(t)=\sum_{ab}L_{ab}e^{i E_{ab}t}\langle a|S|b\rangle$, with $L_{ab}=|a\rangle\langle b|$ and $ E_{ab}=E_a- E_b$, in terms of the eigenbasis of $\tilde{H}_S=\sum_aE_a|a\rangle\langle a|$, and split the terms into those depending on the decaying part of the correlation, $\alpha_B(t)$, and those depending on $C_0$, with $C_B(t)=\alpha_B(t)+C_0$. Then, by assuming a fast decay of $\alpha_B$, such that $\gamma_t(\omega)=\int_0^t d\tau \alpha_B(\tau)e^{i\omega\tau}\approx \int_0^\infty d\tau \alpha_B(\tau)e^{i\omega\tau}$, and also considering the secular approximation, we find (see SM)
 \begin{eqnarray}
\frac{d\rho_s (t)}{dt}
&=&-i\sum_{ab}\Delta_{ab}[L_{ab}L^\dagger_{ab},\rho_s(t)]\cr
&+&\sum_{ab} \gamma_{ab}\bigg[L_{ab}\rho_s(t)L^\dagger_{ab}-\frac{1}{2}\bigg\{L^\dagger_{ab}L_{ab},\rho_s(t)\bigg\}\bigg]\cr
&+&C_0\int_0^t d\tau e^{i E_{ab}\tau} [ L_{ab}\rho_s (\tau),L^\dagger_{ab}]+{\textmd h.c.}\,.
\label{convol3}
\end{eqnarray}
The terms proportional to the offset do not allow the equation to be in a Lindblad form. Thus, as discussed in the SM, the thermal state $\rho_S^\tth\sim \exp(-\beta H_S)$ is no longer a steady state of such equation. Moreover, Eq. (\ref{convol3}) is no longer consistent to its second order time-local counterpart, reflecting a failure of the weak coupling approximation. The convenience of having a decaying-to-zero correlation function for a well-defined weak coupling approximation seems therefore clear. While there might be alternative ways to tackle the problem, all derivations seem to contain assumptions that lead to a zero offset. For instance, Gaspard and Nagaoka \cite{gaspard1999b} derive a weak coupling stochastic Schr\"odinger equation for an open systems coupled to a thermal bath that is valid for arbitrary environments and not only to Gaussian environments like in Refs. \cite{strunz2001,alonso2005}. Importantly, their derivation relies on the assumption of statistical typicality, which ensures a decaying correlation function. 

\textit{Spin-Boson environment--}
To illustrate the emergence of an offset in the environment correlation function, we consider a realistic model consisting of $M$ independent organic dye molecules. Each molecule have a complex structure consisting of two electronic internal states strongly affected by a set of rovibronic modes; thus, it can be treated as a two-party, i.e. hybrid, system that can be well described with the Dicke-Holstein model \cite{kirton2015}:
\begin{eqnarray} 
  H_E&=&\sum_{j}\left\{ H_{\sigma_j}+\sum_\lambda\left(\omega_{j\lambda}b_{j\lambda}^{\dagger}b_{j\lambda}+\sigma^z_j x_{j\lambda}\right)\right\},
  \label{envhamil}
\end{eqnarray}
where the index $j$ runs over all molecules. Here the electronic part is written as: $H_{\sigma_j}=\frac{1}{2}\Omega_j\hat n_j\cdot \vec\sigma_j$, where $\hat n_j=\Omega_j^{-1}(-\Delta_j,0,\varepsilon_j)$ is a unitary vector, and $\vec \sigma_j =(\sigma^x_j,\sigma^y_j,\sigma^z_j)$ is the vector of Pauli matrices. We have defined the Rabi-like frequency $\Omega_j=\sqrt{\Delta_j^2+\varepsilon_j^2}$, with $\Delta_j$ and $\varepsilon_j$ being the spin-tunneling strength and flip-flop energy, respectively. The rovibronic part is represented by the bosonic creation (annihilation) operators $b_{j\lambda}^\dagger$ ($b_{j\lambda}$), with energies $\omega_{j\lambda}$, and reaction coordinate: $x_{j\lambda}=g_{j\lambda}(b_{j\lambda}+b^\dagger_{j\lambda})$, where $g_{j\lambda}$ represents the coupling strength between electronic and vibrational modes. Considering a linear dispersion, this quantity is determined by the spectral function $J(\omega)=(g(\omega))^2$, which we consider Ohmic: $J(\omega) = r^2\frac{\omega}{\omega_c}\exp(-\omega/\omega_c) \theta(2\omega_c-\omega)$ \cite{caldeira1983b}, where $\theta(x)$ is the Heaviside function, $2\omega_c$ the cut-off frequency and $0\leq r\leq 1$ a parameter that controls the spin-boson interaction strength. The molecular environment is in a thermal state with inverse temperature $\beta$. 
 
We consider an open system coupled to such hybrid environment through the standard interaction Hamiltonian $H_I=BS$, where $S$ is an arbitrary system operator and $B= \sum_j\sigma^x_j$, such that 
$C_B(t)$ is a sum of single-molecule correlation functions (SMCF), i.e.: $C_B(t)=\sum_jC^{(j)}_B(t)$, with $C_B^{(j)}(t)={\rm tr}\{\tilde{\sigma}^x_j(t)\tilde{\sigma}^x_j(0)\rho_E^{\rm th}\}$. Thus, since the molecules are not interacting, we only need to study the properties of the $j$th SMCF, $C_B^{(j)}$, which we tackle with exact diagonalization for up to three vibrational modes (in SM) and MPS for more. In the following, we drop the index $j$ for simplicity. 

\textit{Single-Molecule Environment--}
For certain parameter regimes the analysis of the SMCF is quite simple. To show this, we apply the polaron transformation to the single-molecule environment Hamiltonian resulting in: $ H_{\rm SM}=E_{0} +\frac{1}{2}\varepsilon\sigma^z+\sum_\lambda\omega_{\lambda}b^{\dagger}_{\lambda}b_{\lambda}-\frac{1}{2}\Delta (\sigma^+ e^{K}+\sigma^- e^{-K})$, with $E_{0}=\sum_\lambda g_{\lambda}^2/\omega_{\lambda}$, and $\sigma^+$ ($\sigma^-$) being the rising (lowering) spin-$1/2$ operator. Thereby, there exists two limit cases for which the SMCF is analytically accessible:  First, if the spin-tunneling strength $\Delta=0$ the environment Hamiltonian becomes separable, and we obtain the well-known pure dephasing limit. The eigenstates are separable spin-boson bare states: $|\epsilon_k\rangle=|s,{\bf n}\rangle$, with $s=0,1$ labeling the electronic state of the molecule, and $|{\bf n}\rangle\equiv|n_{1}\cdots n_{L}\rangle$ its respective multimode Fock state . Accordingly, the offset $C_0=0$, since $B^{kk}=\langle \epsilon_k|\sigma^x|\epsilon_k\rangle=0$, for all $k$. Furthermore, after some analytics, it is possible to prove that $C_B(t)\propto e^{-\Gamma_{\beta}(t)}$, where $\Gamma_\beta(t) = 8\sum_\lambda (g_{\lambda}/\omega_{\lambda})^2\sin^2(\omega_{\lambda} t)\coth(\beta\omega_{\lambda}/2)$ (see SM).

The second case is when $r\approx 0$, which corresponds to a spin environment with negligible vibrational part. $H_{\rm SM}$ is again separable and, as shown in the SM, the SMCF reduces to $C_B(t) = (\varepsilon/\Omega)^2 \cos(\Omega t)+C_0$, with the offset given by $C_0=(\Delta/\Omega)^2{\rm sech}^2(\beta\Omega/2)$. Thus, a non-zero offset requires that $\Delta\neq 0$ and a finite temperature $\beta^{-1}> 0$. Nonetheless, the offset disappears at zero temperatures even when $r\neq 0$. This is shown by rewriting the thermal weights of $\rho_E^\tth$ as: $\eta_{k}=e^{-\beta\epsilon_k}/\sum_ke^{-\beta\epsilon_k} = e^{-\beta(\epsilon_k-\epsilon_{0})}/(1+\sum_{k\neq 0}e^{-\beta(\epsilon_k-\epsilon_{0})})$, where $\epsilon_{0}<\epsilon_k$ is the ground state. By setting $\beta\rightarrow \infty$ we find that $\eta_{k}\rightarrow 1$, iff, $\epsilon_k=\epsilon_{0}$, and zero otherwise. Thus, in the zero temperature limit, the only eigenstate involved in Eq.~\ref{offset} is the ground state, such that the condition $(b)$ is no longer fulfilled, and therefore $C_0=0$.


Away from the above limits the dynamics of the environment displays a competition between the dephasing and the spin-coherence. While the offset appears when $\Delta\neq 0$, resulting in an oscillatory behavior of the SMCF, the role of $r$ is to induce a damping. In this scenario the eigenstate structure of the Hamiltonian is analytically and numerically hard to access, particularly when dealing with many vibrational modes. However, since any eigenstate can be written as linear combination of the spin-boson bare states, i.e.,  $|\epsilon_k\rangle=\sum_{s{\bf n}}c^{(k)}_{s{\bf n}} |s,{\bf n}\rangle$, we can infer that a minimal requirement for the offset is that the eigenstates display a mixing of at least two spin states, that is,  $B^{kk}=\langle \epsilon_k|\sigma^x|\epsilon_k\rangle= \sum_{\bf n}(c^{(k)}_{1{\bf n}}(c^{(k)}_{0{\bf n}})^* \langle 0|\sigma^x|1\rangle+c^{(k)}_{0{\bf n}}(c^{(k)}_{1{\bf n}})^* \langle 1|\sigma^x|0\rangle)\neq 0$. While spin coherences are required for a finite offset, certain entangled eigenstates may have zero contribution. For instance, an entangled eigenstate of the form $|\epsilon_k\rangle= c^{(k)}_{0,{\bf n}}|0,{\bf n}\rangle+c^{(k)}_{1{\bf m}}|1,{\bf m}\rangle$, with $|{\bf n}\rangle \neq |{\bf m}\rangle$, leads to $\langle \epsilon_k|\sigma^x|\epsilon_k\rangle =0$.

\begin{figure}[ht]
\includegraphics[width=\columnwidth]{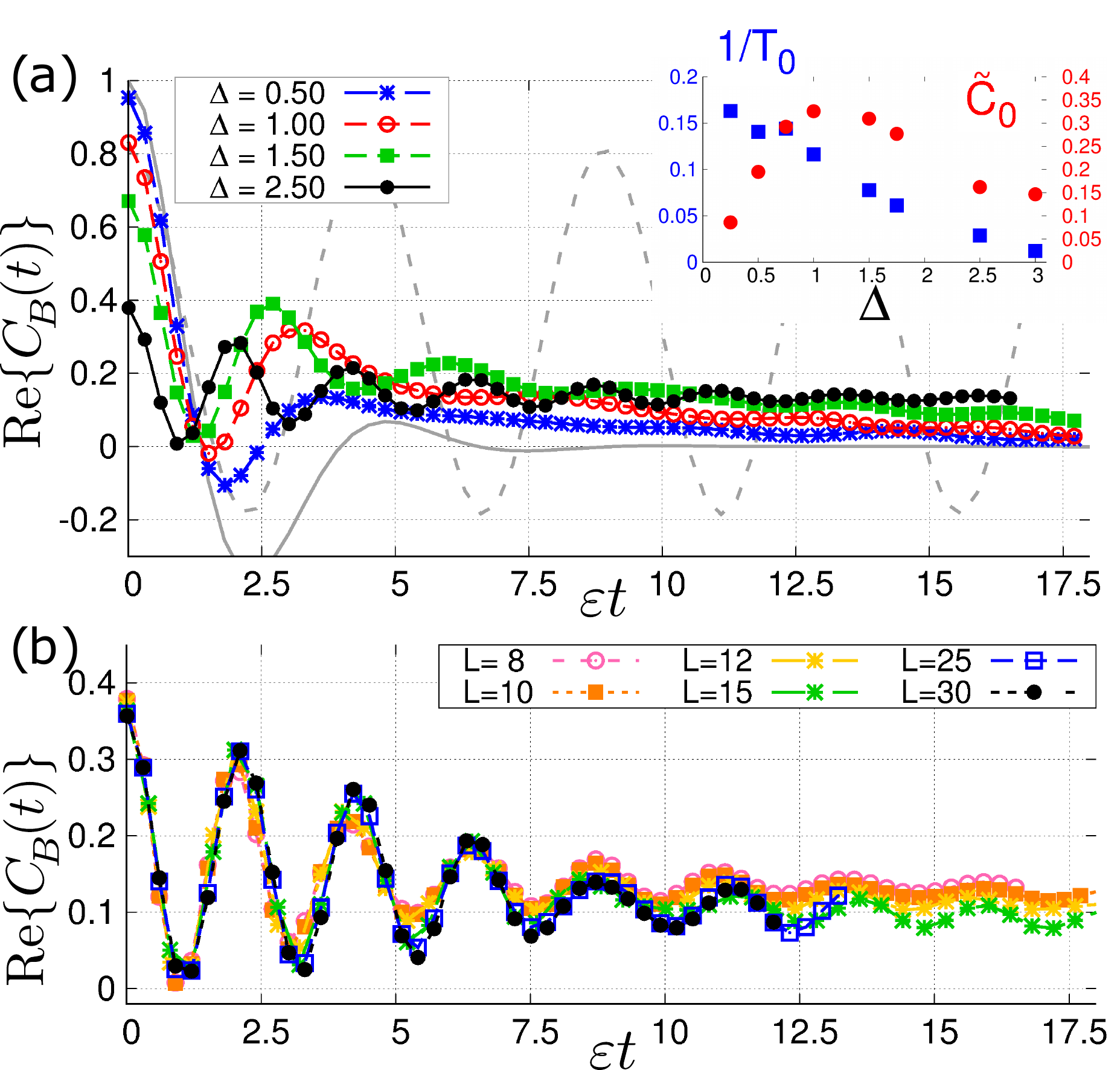} 
\caption{\label{fig:01} {\bf SMCF:}  (a) Real part of the MPS-computed SMCF for different values of $\Delta$ with $r=0.25$ for eight-mode ($L=8$) dye-molecule. The light gray lines are the related trivial limits: spin-coherence (dashed) $r=0$ and pure-dephasing (no dashed) $\Delta=0$. The inset shows correlation time $T_0$ (blue filled squares) and the pre-offset term $\tilde C_0$ (red filled circles) as a function of $\Delta$. (b) Scaling of the correlation function with the number of bosonic modes $L$ for $r=0.25 \varepsilon$ and $\Delta=2.5 \varepsilon$. These calculation were done for $\beta=1$ and local MPS dimension $d_{\rm MPS}=10...15$ which showed to be sufficient for good convergence of our results.}
\end{figure}

In general, the coherence-dephasing competition can be qualitatively characterized in terms of the ratio $r/\Delta$. For intermediate values of $r/\Delta$, we consider the matrix product state (MPS) formalism \cite{scholl2011,perez2007} implemented in the MPS library \cite{hubig:_syten_toolk,hubig17:_symmet_protec_tensor_networ} to compute the SMCF. This is shown in Fig.~\ref{fig:01} for different values of $\Delta$ (upper panel) and different numbers $L$ of bosonic modes (lower panel), which shows that the offset does not depend on the environment size.
In this regard, our analysis (see details in the SM) suggests that the best fitting function for the real part of the SMCF in Fig.~\ref{fig:01} is 
\bea
f(t)=A_0 \cos(\omega_0 t) e^{-B_0 t^a}+\tilde C_0 e^{-t/T_0}. 
\label{fit}
\eea
Thus, the short time non-Markovian decay is dominated by the first factor, which represents a non-exponential decay since $a\approx \sqrt{2}>1$, while the long time limit dynamics is dominated by an exponential decay with a characteristic -correlation- time $T_0$ that takes different values. In the inset of Fig.~\ref{fig:01}-(a) we present the relevant fitting parameters, $1/T_0$ and $\tilde C_0$, as a function of $\Delta$. Towards the limit $r/\Delta \rightarrow 0$,  the correlation time $T_0$ goes to infinity, while the prefactor defines the resulting offset $\tilde C_0\rightarrow  C_0$. 

{\emph{Many-Molecule Environment--}}
Finally, when considering the case of $M$ molecules the total offset is enhanced since $C_0=\sum_jC^{(j)}_0$. Moreover, the ensemble of molecules behave at long times as a collection of exponentially decaying two-level systems, each having different internal parameters and thus different decay rates $1/T_0$. As it is well-known this type of systems leads to a $1/f$ spectrum between the frequency ranges of $[1/T_{\textmd{max}},1/T_{\textmd{min}}]$, where $T_{\textmd{max}} (T_{\textmd{min}})$ are the maximum (minimum) correlation times \cite{paladino2014,milotti2002,faoro2004}. Interestingly, contrary to the standard case where no-offset is considered, here we find that for some molecule $T_0=T_{\textmd{max}}\sim \infty$, which implies that the $1/f$ extends towards the zero frequency limit, as it is observed experimentally \cite{milotti2002} (see SM).

\textit{Conclusions and outlook--}
We argue that a necessary condition for the existence of a well-defined weak coupling approximation is to have a correlation function that decays to zero in time, a feature that is fulfilled for Gaussian and ETH environments but not for general thermal environments. To illustrate this we consider a hybrid spin-boson environment that presents a non-decaying correlation function, and show how this connects with the presence of a 1/f spectrum that, in contrast to previous analysis, extends to zero frequencies. This work opens several interesting research avenues. Firstly, quantum devices such as superconducting qubits are affected by similar hybrid environments \cite{norris2016,sung2019} and 1/f noise \cite{paladino2014,milotti2002,faoro2004}, and thus a rigorous analysis of the convergence of a weak coupling expansion (and the related Lindblad equation) may be important. Secondly, this work may be helpful to understand the dynamics of impurities coupled to materials that do not fulfil ETH, for instance those presenting dynamical localization or Hilbert space fragmentation, where the emergence of an offset has also been described \cite{sala2019,khemani2019,alhambra2019}.
Further, the offset is linked to strong memory effects and to the inability of the environment to bounce back to its equilibrium state after interacting with the system, which raises an interest in characterizing its non-Markovianity  \cite{rivas2009,breuer2009,hall2014}. Finally, due to the non-ergodicity of the environment, the system quantum information may not be irreversibly lost and could be recovered by developing appropriate correction protocols. 

\begin{acknowledgements}
The authors gratefully acknowledge A. Recati, U. Schollw\"ock, L. Viola, D. Wierichs and B. A. Rodriguez for interesting discussions. IDV and CP are financially supported by DFG-grant GZ: VE 993/1-1. CH acknowledges the support from ERC Grant QUENOCOBA, ERC-2016-ADG (Grant No. 742102).
\end{acknowledgements}
\bibliography{/Users/ines.vega/Dropbox/Bibtexelesdrop2}

\appendix

\section{Appendix A: Master equation for correlation functions with an offset}\label{App1}
Up to second order in the weak coupling parameter, and in interaction picture with respect to $\tilde{H}_S$, the master equation reads as \cite{breuerbook}
\begin{eqnarray}
\frac{d\rho_s (t)}{dt}=
\int_0^t d\tau C_B(t-\tau) [ S(\tau) \rho_s (t-\tau),S(t)]+{\textmd h.c.}\nonumber
\label{convol1}
\end{eqnarray}
where we have crucially considered the Born approximation, by which within the right hand side term one can reply the total density operator as $\rho(t-\tau)\approx \rho_s(t-\tau)\otimes\rho_E$. Moreover, we have defined the environment correlation function as Eq. (\ref{eq:2}). The above equation can now be separated in two terms
\begin{eqnarray}
&&\frac{d\rho_s (t)}{dt}=
\int_0^t d\tau \sum_{lj} \alpha_B(\tau) [ S(t-\tau) \rho_s (t),S(t)]+{\textmd h.c.}\cr
&+&C_0\int_0^t d\tau \sum_{lj}  [ S(t-\tau) \rho_s (\tau),S(t)]+{\textmd h.c.},
\label{convol2}
\end{eqnarray}
where we have separated $C_B(t)=\alpha_B(t)+C_0$ in terms of the time dependent part and the offset. Moreover, we assume that $\alpha_B(t)$ decays very fast, so that it is a good approximation to consider $\rho(\tau)\approx \rho(t)$. Nevertheless, such replacement can not be made in the term that depends on the offset, where the integrand is not ensured to have a fast decaying. We now assume that 
\bea
S(t)=\sum_{ab}L_{ab}e^{i E_{ab}t}\langle a|S|b\rangle
\eea
where $L_{ab}=|a\rangle\langle b|$, in terms of the eigenstates and eigenvalues of $\tilde{H}_S$. Therefore, we find that 
 \begin{eqnarray}
&&\frac{d\rho_s (t)}{dt}
=-i\sum_{ab,cd}\Delta^t_{ab,cd}[L_{ab}L^\dagger_{cd},\rho_s(t)]\\
&+&\sum_{ab,cd} \gamma^t_{ab,cd}\bigg[L_{ab}\rho_s(t)L^\dagger_{cd}-\frac{1}{2}\bigg\{L^\dagger_{cd}L_{ab},\rho_s(t)\bigg\}\bigg]\cr
&+&C_0 e^{i(E_{ba}-E_{dc})t}\int_0^t d\tau e^{i E_{ab}\tau} [ L_{ab}\rho_s (\tau),L^\dagger_{cd}]+{\textmd h.c.}\nonumber
\label{convol3a}
\end{eqnarray}
with $E_{ba}= E_b-E_{a}$ and the coefficients defined as 
\bea
\gamma^t_{ab,cd}&=&e^{i(E_{ba}-E_{dc})t}\gamma(E_{ba}) \langle a|S|b\rangle\langle d|S |c\rangle,\cr
\Delta^t_{ab,cd}&=&e^{i(E_{ba}-E_{dc})t}\Sigma(E_{ba})\langle a|S|b\rangle\langle d|S |c\rangle,.
\label{gammadelta}
\eea 
where we have defined 
\bea
\gamma(\omega)&=&\Rre\bigg\{\int_{0}^\infty \alpha_B(\tau) e^{i\omega \tau}d\tau\bigg\} \cr
\Sigma(\omega)&=&\Iim\bigg\{\int_{0}^\infty \alpha_B(\tau) e^{i\omega \tau}d\tau\bigg\}
\label{terms}
\eea
In this point, we realize that up to the second order considered the convoluted Eq. (\ref{convol3a}) should be equivalent to its time-local counterpart. This is because the terms on its r.h.s are already of second order, and therefore we can make the replacement $\rho_s(\tau)=\rho_s(t)+{\mathcal O}(g^2)$ to find
 \begin{eqnarray}
&&\frac{d\rho_s (t)}{dt}
=-i\sum_{ab,cd}\Delta^t_{ab,cd}[L_{ab}L^\dagger_{cd},\rho_s(t)]\\
&+&\sum_{ab,cd} \gamma^t_{ab,cd}\bigg[L_{ab}\rho_s(t)L^\dagger_{cd}-\frac{1}{2}\bigg\{L^\dagger_{cd}L_{ab},\rho_s(t)\bigg\}\bigg]\cr
&+&C_0 e^{i(E_{ba}-E_{dc})t}\int_0^t d\tau e^{i E_{ab}\tau} [ L_{ab}\rho_s (t),L^\dagger_{cd}]+{\textmd h.c.}\nonumber
\label{timelocal}
\end{eqnarray}
Nonetheless, if the offset is large the two equations will differ significantly, signaling the failure of the weak coupling. Indeed, when the offset is large the time local master equation (\ref{timelocal}) becomes increasingly unstable, as the coefficients related to $C_0$ grow unbounded. In contrast, the convoluted equation appears to be more stable but reflects a dependency or memory over the whole trajectory of $\rho_s(\tau)$ that suggest a strong non-Markovian character and thus compromises the weak coupling assumption. To see this we proceed further, and consider the standard secular approximation in Eq. (\ref{convol3a}), to find 
 \begin{eqnarray}
&&\frac{d\rho_s (t)}{dt}
=-i[H_{\textmd{eff}},\rho_s(t)]\cr
&+&\sum_{ab} \gamma_{ab}\bigg[L_{ab}\rho_s(t)L^\dagger_{ab}-\frac{1}{2}\bigg\{L^\dagger_{ab}L_{ab},\rho_s(t)\bigg\}\bigg]\cr
&+&C_0\int_0^t d\tau e^{i E_{ab}\tau} [ L_{ab}\rho_s (\tau),L^\dagger_{ab}]+{\textmd h.c.}
\label{convol4}
\end{eqnarray}
where we have defined $H_{\textmd{eff}}=\sum_{ab}\Delta_{ab}L_{ab}L^\dagger_{ab}$, $\gamma_{ab}=\gamma(E_{ba}) \langle a|S|b\rangle\langle b|S |a\rangle$ and $\Delta_{ab}=\Sigma(E_{ba})\langle a|S|b\rangle\langle b|S |a\rangle$.
In the long time limit we formally have 
\bea
&&\frac{d\rho_s (t)}{dt}
=-i[H_{\textmd{eff}},\rho_s(t)]
+\sum_{ab} \gamma_{ab}\bigg[L_{ab}\rho_s(t)L^\dagger_{ab}\cr
&-&\frac{1}{2}\bigg\{L^\dagger_{ab}L_{ab},\rho_s(t)\bigg\}\bigg]+C_0\sum_{ab} \bigg[L_{ab}\rho_s(t,E_{a}- E_b)L^\dagger_{ab}\cr
&-&\frac{1}{2}\bigg\{L^\dagger_{ab}L_{ab},\rho_s(t,E_{a}- E_b)\bigg\}\bigg],
\label{lindblad_offset}
\eea
which depends on the quantity 
\bea
\rho_s(t,\omega)=\int_0^t d\tau e^{i\omega \tau}\rho_s(\tau).
\eea
When projected into the system eigenbasis, one can calculate the rate equations as before, 
\bea
&&\frac{d\langle m|\rho_s(t)|m\rangle}{dt}=\sum_{b\neq m}\bigg(\gamma_{mb}\langle b|\rho_s(t)|b\rangle\cr
&-&\gamma_{bm}\langle m|\rho_s(t)|m\rangle\bigg)
+C_0\bigg(\sum_{b\neq m}\langle b|\rho_s(t,E_m- E_b)|b\rangle\cr
&-&\langle m|\rho_s(t,E_m- E_b)|m\rangle\bigg)
\eea
Thus, the steady state condition is 
\bea
&&\sum_{b\neq m}\bigg(\gamma_{mb}P^{\sst}_{bb}-\gamma_{bm}P^{\sst}_{mm}\bigg)=\cr
&-&C_0\sum_{b\neq m}\bigg(\langle b|\rho_s(t\rightarrow\infty,E_m- E_b)|b\rangle\cr
&-&\langle m|\rho_s(t\rightarrow\infty,E_m- E_b)|m\rangle\bigg).
\eea
In the standard case, we have that $C_0=0$ and thus, one steady state solution is the thermal one, i.e. $P^{\sst}_{bb}\sim\exp(-\beta E_b)$ provided that the detailed balance is also fulfilled, i.e. $\gamma_{mb}=\exp(-\beta E_{mb})\gamma_{bm}$. Nonetheless, when $C_0\neq 0$ we find that $P^{\sst}_{bb}$ is no longer a thermal state, since is given by 
\bea
&&\gamma_{mb}P^{\sst}_{bb}-\gamma_{bm}P^{\sst}_{mm}=-C_0\bigg(\langle b|\rho_s(t\rightarrow\infty,E_m- E_b)|b\rangle\cr
&-&\langle m|\rho_s(t\rightarrow\infty,E_m- E_b)|m\rangle\bigg),
\eea
and thus it depends on the whole history of evolution, including the initial state.

\section{Appendix B: Environment in extreme cases}\label{App2}
Here we shortly present the derivation of the formulae used in the main text for the cases: (a) pure-dephasing $\Delta=0$ and (b) spin-coherence, $r=0$. In addition, we also present a case (c) corresponding to weak-coupling derivation, that will allow us to justify in this limit the fitting functional (\ref{fit}) used to characterize the single-molecule correlation function (SMCF). First, we stress that throughout the paper we have only computed single-molecule correlation corresponding to the $j$th molecule due to the non-direct interaction between them. Hence the full environment Hamiltonian can be written as $H_E=\sum_jH_E^{(j)}$ and the many-molecule  polaron transformation gets defined then as $U_{\rm full}=U_{(1)}\otimes\cdots\otimes U_{(j)}\otimes\cdots\otimes U_{(N_{\rm mol})}$, where
\begin{eqnarray}
  U_{(j)}=\exp(\frac{1}{2}\sigma^z_j K_j)\,,
\end{eqnarray}
with $K_j= 2\sum_\lambda \frac{g_{j\lambda}}{\omega_{j\lambda}}(b^\dagger_{j\lambda}-b_{j\lambda})$.  Henceforth, in the computation of the SMCF we ignore the index $j$, therefore we redefine $H^{(j)}_E\equiv H_{\rm SM}$. Considering that the formula for the SMCF implies the computation of a trace, we can apply arbitrary unitary transformation $U$
with which we can write
\begin{eqnarray}
  C_B(t)&=&{\rm tr}_E\left\{e^{-i t H_{\rm SM}}\sigma^xe^{i t H_{\rm SM}}\sigma^x\rho^{\rm th}_{E}(0)\right\}\nonumber\\
  &=&{\rm tr}_E\left\{e^{-i t \overline H_{\rm SM}}\overline \sigma^xe^{i t \overline H_{\rm SM}}\overline\sigma^x\overline\rho^{\rm th}_{E}(0)\right\}\nonumber\\
\end{eqnarray}
where
\begin{eqnarray}
  U e^{-i t H_{\rm SM}} U^{-1}&=&\sum_p\frac{(-i t)^p}{p!}U \underbrace{H_{\rm SM}\cdots H_{\rm SM}}_\text{$p$ times} U^{-1}\nonumber\\
  &=&\sum_p\frac{(-i t)^p}{p!}\underbrace{\overline H_{\rm SM}\cdots \overline H_{\rm SM}}_\text{$p$ times}\nonumber\\
&=&e^{-i t \overline H_{\rm SM}}\,,
\end{eqnarray}
where we have used the short notation $\overline H_{\rm SM}=U H_{\rm SM} U^{-1}$. It can be also shown that
\begin{eqnarray*}
  \overline\sigma^x &=& U\sigma^x U^{-1} = \sigma^+ e^{K}+\sigma^- e^{-K},\\
  \overline\sigma^z &=& U\sigma^x U^{-1} = \sigma^z,\\
  \overline b_{\lambda} &=& U b_{\lambda} U^{-1}= b_{\lambda}-\frac{g_{\lambda}}{\omega_{\lambda}}\sigma^z,
\end{eqnarray*}
with which $H_{\rm SM}$ gets transformed into:
\begin{eqnarray}
  \overline H_{\rm SM} &=&\frac{\varepsilon}{2}\sigma^z+\sum_\lambda \omega_\lambda b^\dagger_{\lambda} b_{\lambda}+\sum_{\lambda}\frac{g^2_{\lambda}}{\omega_{\lambda}}\nonumber\\
  && -\frac{\Delta}{2}(\sigma^+ e^{K}+\sigma^- e^{-K}).\,
\end{eqnarray}

\subsection{(a) Pure dephasing case } Let us first consider $\Delta=0$ and this Hamiltonian is separable, and thus $e^{-i t \overline H_{\rm SM}}=e^{-\frac{1}{2}it\varepsilon\sigma^z} e^{-i t \sum\omega_{\lambda} b^\dagger_{\lambda} b_{\lambda}}$. Then, the SMCF can be expressed as
\begin{eqnarray}
  C_B(t)&=&\langle \overline\sigma^x (t)\overline\sigma^x(0)\rangle_{E}-
  \langle\overline{\sigma}^x\rangle_E^2\nonumber\\
  &=&e^{i\varepsilon t}\langle \sigma^+\sigma^-\rangle_{\sigma}\langle e^{K(t)}e^{-K(0)}\rangle_R\nonumber\\
  &{}&+e^{-i\varepsilon t}\langle \sigma^-\sigma^+\rangle_\sigma\langle e^{-K(t)}e^{K(0)}\rangle_R\nonumber\\
  &{}&-\langle \sigma^+\rangle_\sigma\langle e^{K(0)} \rangle_R-\langle \sigma^-\rangle_\sigma\langle e^{-K(0)} \rangle_R\,,
\end{eqnarray}
where we have used the separability of $\overline H_{\rm SM}$ to write $\rho_{E}^{\rm th}(0)=\rho_\sigma\otimes \rho_R$, with $\rho_\sigma=e^{-\frac{1}{2}\beta\varepsilon \sigma^z}/Z_\sigma$ and $\rho_R=e^{-\beta\sum_\lambda\omega_{\lambda} b^\dagger_{\lambda} b_{\lambda}}/Z_R$.  Thus, the respective traces are denoted as $\ttr_\sigma\{\cdots\rho_\sigma\}=\langle\cdots\rangle_\sigma$, and $\ttr_\sigma\{\cdots\rho_R\}=\langle\cdots\rangle_R$. Further, the operator $K(x)$ is defined as
\begin{eqnarray}
  K(x)&=&e^{i x\sum_\lambda \omega_{\lambda} b^\dagger_{\lambda} b_{\lambda}} K e^{-i x\sum_\lambda\omega_{\lambda} b^\dagger_{\lambda} b_{\lambda}}\nonumber\\
  &=&2\sum_\lambda \frac{g_{\lambda}^2}{\omega_{\lambda}^2} (b^\dagger_{\lambda} e^{ix}-b_{\lambda} e^{-i x})\,,
\end{eqnarray}
that satisfies
\begin{eqnarray}
  [K(x),[K(x),K(y)]]=[K(y),[K(x),K(y)]]=0,\nonumber
\end{eqnarray}
and
\begin{eqnarray}
  [K(x),K(y)]=i8 \sum_\lambda \frac{g_{\lambda}^2}{\omega_{\lambda}^2}\sin(\omega_{\lambda}(x-y)).
\end{eqnarray}
 Therefore, we can use the Baker-Kambell-Hausdorf identity 
\begin{equation}
e^{K(x)}e^{-K(y)}=e^{K(x)-K(y)-\frac{1}{2}[K(x),K(y)]}\,,
\end{equation}
and the thermal average
\begin{eqnarray}
\langle e^{\sum_\lambda(c_\lambda b_\lambda+d_\lambda b^{\dagger}_\lambda)}\rangle_R=e^{\sum_\lambda c_\lambda d_\lambda (2 n(\omega_\lambda)+1)}
\end{eqnarray}
with the thermal distribution $n(\omega_\lambda)=(e^{\beta\omega_\lambda}-1)^{-1}$, and the complex numbers $c_\lambda$ and $d_\lambda$. With these formulae, we can compute the two-point thermal correlator of the type $\langle e^{K(x)}e^{-K(y)}\rangle_R=\exp(\sum_\lambda\frac{g^2_{\lambda}}{\omega^2_{\lambda}} h_{\lambda}(\beta))$ with
\begin{eqnarray}
h_{\lambda}(\beta)=-2|\xi_{\lambda}(x,y)|^2 \coth(\beta\omega_{\lambda}/2)+i 4 \mu_{\lambda}(x,y)
\end{eqnarray}
with
\begin{eqnarray*}
  \xi_{\lambda}(x,y)&=&i 2 e^{i\omega_{\lambda}(x+y)/2}\sin[\omega_{\lambda}(x-y)/2]\,\\
  \mu_{\lambda}(x,y)&=& \sin[\omega_{\lambda}(x-y)]\,.
\end{eqnarray*}
Considering this, together with the identity $\langle e^{K(x)}e^{-K(y)}\rangle_R=\langle e^{K(y)}e^{-K(x)}\rangle^*_R$, and that for a thermal initial state $\langle \sigma^\pm\rangle_\sigma=0$ and $\langle \sigma^+\sigma^-\rangle_\sigma=(1+e^{-\beta\varepsilon})^{-1}$, we obtain the final expression for the SMCF:
\begin{eqnarray}
  C_B(t) &=& \left(\frac{e^{i\varepsilon t}}{1+e^{-\beta\varepsilon}}+\frac{e^{-i\varepsilon t}}{1+e^{\beta\varepsilon}}\right)\nonumber\\
  & &\times\exp\left(-i 4 \sum_{\lambda} \frac{g_{\lambda}^2}{\omega_{\lambda}^2}\sin(\omega_{\lambda} t)\right)\nonumber\\
  & &\times\exp\left(-8\sum_{\lambda} \frac{g_{\lambda}^2}{\omega_{\lambda}^2}\sin^2(\omega_{\lambda} t)\coth(\beta\omega_{\lambda}/2)\right),\nonumber\\
\end{eqnarray}
which implies no offset, as expected, for the pure-dephasing case.
 
 \subsection{(b) No dissipation}
 Let us now consider $r\approx 0$. The polaron-transformed Hamiltonian $H_{\rm SM}$ takes the the form
 \begin{eqnarray}
   \overline H_{\rm SM}=\frac{1}{2}\Omega\, \hat n \cdot \vec\sigma +\sum_\lambda \omega_{\lambda} b^\dagger_{\lambda} b_{\lambda}\,,
 \end{eqnarray}
  since $g_{\lambda} \propto r \approx 0\Rightarrow e^{\pm K}\approx 1$. Here we have defined a Rabi-type of frequency $\Omega=\sqrt{\Delta^2+\varepsilon^2}$,  $\hat n = \Omega^{-1}\left(-\Delta,0,\varepsilon\right)$ is a unitary operator, and $\vec\sigma=(\sigma^x,\sigma^y,\sigma^z)$. To compute the SMCF, we directly use the identity
\begin{eqnarray}\label{spinid}
 e^{i\alpha \hat n\cdot {\vec \sigma}}=\hat 1 \cos(\alpha)+i(\hat n\cdot {\vec \sigma}) \sin(\alpha)\,,
\end{eqnarray}
with $\alpha=\Omega_j t/2,$ and the separability of $\overline H_{\rm SM}$ to express:
\begin{eqnarray}
  \sigma^x(t)\sigma^x(0) &=& e^{i t\overline H_{\rm SM}}\sigma^x e^{-i t \overline H_{\rm SM}}\sigma^x\nonumber\\
  &=&\cos^2(\Omega t/2)+\frac{\Delta^2-\varepsilon^2}{\Delta^2+\varepsilon^2}\sin^2(\Omega t/2)\nonumber\\
  & & -2\frac{\Delta \varepsilon}{\Delta^2+\varepsilon^2}\sigma^z\sigma^x\sin^2(\Omega t/2)\,,
\end{eqnarray}
which after computing the thermal average $\langle\cdots\rangle_E|_{r=0}$ results in:
\begin{eqnarray}
\langle \sigma^x(t)\sigma^x(0)\rangle_E|_{r=0}=\frac{1}{\Omega}\left(\Delta^2-\varepsilon^2+2\varepsilon^2\cos^2(\Omega t/2)\right),\nonumber\\
\end{eqnarray}
since $\langle\sigma^z\sigma^x\rangle_E|_{r=0}=0$. For the renormalization term we have:
\begin{eqnarray}
  \langle \sigma^x\rangle_E |_{r=0} &=& {\rm tr}_\sigma\{\sigma^x\rho_{\sigma}\}{\rm tr}_{R}\{\rho_R\}\nonumber\\
  &=&\frac{1}{Z_B}{\rm tr}\{\sigma^x\exp(-\beta \hat n\cdot \vec\sigma)\}\nonumber\\
  &=& -\frac{\Delta}{\Omega}{\rm \tanh}(\beta\Omega/2)
\end{eqnarray}
where we have again use the identity~(\ref{spinid}) with $i\alpha=-\beta \Omega/2$, and with which the partition function can be written as $Z_{\sigma}=\cosh(\beta\Omega/2)$. We finally arrive to the SMCF:
\begin{eqnarray}
  C_B(t)&=&\langle \sigma^x(t)\sigma^x(0)\rangle_E|_{r=0}-(\langle\sigma^x\rangle_E|_{r=0})^2\nonumber\\
  &=&\frac{\varepsilon^2}{\Omega^2}\cos(\Omega t)+\frac{\Delta^2_j}{\Omega^2}{\rm sech}^2(\beta\Omega/2)\,,
\end{eqnarray}
where we have the trigonometrical identities: $\cos(2\theta)=2\cos^2(\theta)-1$, and ${\rm sech}^2(x)=1-\tanh^2(x)$. Clearly the last term in the right  side corresponds to the offset, that is
\begin{eqnarray}
C_0=\frac{\Delta^2}{\Omega^2}{\rm sech}^2(\beta\Omega/2)\,.
\end{eqnarray}

\subsection{(c) Weak coupling also for the initial state}
 
Another possibility to approach the SMCF is to consider the case in which the interaction between the electronic and vibrational parts of $H_E$ is weak, such that the initial thermal state can be approximated as 
\bea
\rho_E^\tth= \rho^\tth_\sigma\otimes \rho^\tth_R+\rho_{\textmd{corr}}\approx \rho_\sigma^\tth\otimes \rho^\tth_R,
\label{separable}
\eea 
where we consider that $\rho_\sigma^\tth=\exp(-\beta H_{\sigma})/Z_\sigma$ and $\rho^\tth_R=\exp(-\beta H_R)/Z_R$, with $Z_{\sigma,R}$ the partition functions, and $H_R=\sum_\lambda\omega_{\lambda }b_{\lambda}^\dagger b_{\lambda}$.
Thus, we are neglecting the correlation term $\rho_{\textmd{corr}}$, which is of order $r^2$. Moreover, not only this approximation restricts the results to a weak electron-boson coupling; but also it leads to an environment initial state that is no longer in equilibrium, which means that strictly speaking the correlation function is no longer stationary, i.e. $C_B(t_1,t_2)\neq C(t_1-t_2)$. 
For simplicity, we concentrate in the case $t_1=t$ and $t_2=0$. Using the cyclic property of the trace, we write the correlation function as follows 
\begin{equation}\label{weakcoup}
\langle \sigma^x(t)\sigma^x(0)\rangle_E={\rm tr}_\sigma\left\{\sigma^x\tilde \rho_\sigma(t)\right\}\,,
\end{equation}
where $\tilde\rho_\sigma(t)={\rm tr}_R\left\{e^{-i t H_{\rm SM}}\sigma^x\rho^\tth_\sigma\otimes\rho^\tth_Re^{i t H_{\rm SM}}\right\}$. Therefore, we can use the eigenstates of the free spin part of the Hamiltonian $H_\sigma=\frac{1}{2}(\varepsilon\sigma^z-\Delta\sigma^x)$, labeled by $|\mu\rangle$ ($\mu=\pm$), to express Eq.~\ref{weakcoup} as
\begin{equation}
\langle \sigma^x(t)\sigma^x(0)\rangle_E=\sum_{\mu\nu}\langle \mu|\sigma^x|\nu\rangle \tilde\rho_{\nu\mu}(t)\,,
\end{equation}
with $\tilde\rho_{\nu\mu}(t)=\langle \nu|\tilde\rho_\sigma(t)|\mu\rangle$. Following \cite{schaller2015} for instance, we can obtain the matrix element of $\tilde\rho_\sigma$ by solving the following sytem of differential equations
\begin{eqnarray}
\frac{\partial}{\partial t}\tilde\rho_{++} &=& \gamma_{(+-,+-)}\tilde\rho_{++}-\gamma_{(-+,-+)}\tilde\rho_{--}\nonumber\\
\frac{\partial}{\partial t}\tilde\rho_{--} &=& \gamma_{(-+,-+)}\tilde\rho_{--}-\gamma_{(+-,+-)}\tilde\rho_{++}\nonumber\\
\frac{\partial}{\partial t}\tilde\rho_{-+} &=& -i(\varepsilon_--\varepsilon_+ + \sigma_{--}-\sigma_{++})\tilde\rho_{-+}\nonumber\\
&&+\left(\gamma_{(--,++)}-\frac{\gamma_{(+-,-+)}+\gamma_{(+-,+-)}}{2}\right)\tilde\rho_{-+}\,,\nonumber\\
\end{eqnarray}
where $\gamma_{(\mu\nu,\mu'\nu')}$ are relevant damping coefficients and $\sigma_{\pm,\pm}$ are the Lamb-shift terms. The initial conditions are set by $\tilde\rho_\sigma(0)={\rm tr}_\sigma\{\sigma^x\rho_\sigma(0)\}$.
The solutions of the system of diffferential equation have the form:
\begin{eqnarray}
  \tilde\rho_{++}(t)&=&a_1+b_1 e^{-\lambda t},\nonumber\\
  \tilde\rho_{--}(t)&=&a_2+b_2 e^{-\lambda t},\nonumber\\
  \tilde\rho_{+-}(t)&=&b_3 e^{-\lambda_0 t}\,,
\end{eqnarray}
where $\lambda=\gamma_{(-+,-+)}+\gamma_{(+-,+-)}$. The $a_j$ and $b_j$ ($j=1,2$) are real, while $\lambda_0$ and  $b_3$ are complex, and all of them depend explicitly on the damping and the Lamb-shift coefficients, as well as on the initial state $\tilde\rho_{\mu\nu}(0)$. The non-renormalized SMCF is then given by the function
\begin{eqnarray}
  \langle \sigma^x(t)\sigma^x_j(0)\rangle_E&=&\langle -|\sigma^x|-\rangle a_1+\langle +|\sigma^x|+\rangle  a_2 \\
  &+& e^{-\lambda t} (\langle -|\sigma^x|-\rangle b_1+\langle +|\sigma^x|+\rangle  b_2)\nonumber\\
  &+& \langle -|\sigma^x|+\rangle b_3 e^{-\lambda_0 t} +\langle +|\sigma^x|-\rangle b_3^* e^{-\lambda_0^* t}\,.\nonumber
\end{eqnarray}
Interestingly, the form of this solution is very similar to the one of the fitting functional (\ref{fit}) in the main text. Indeed, once renormalized, i.e. after substracting $\langle\sigma^x\rangle^2$, its real part takes the form
\begin{eqnarray}\label{weakfun}
  f^{\textmd{r$\ll$1}}(t) &=&{\rm Re}\{\langle \sigma^x(t)\sigma^x(0)\rangle_E\} -\langle\sigma^x\rangle^2\nonumber\\
  &=&  \tilde A_0 e^{-{\rm Re}\{\lambda_0\} t} \cos({\rm Im}\{\lambda_0\} t)+C_0 e^{-\lambda t}\nonumber\\
  &&+b_0-\langle\sigma^x\rangle^2\,.
\end{eqnarray}
The first term at the right-hand-side corresponds to an exponentially decaying damping $\sim e^{-{\rm Re}\{\lambda_0\} t}$. This is due to the fact that we are assuming the weak coupling approximation between electrons and phonons, while the analogous term in the fitting function (\ref{fit}) presents a non-Markovian structure, $\sim e^{- B_0 t^a}$ with $a\approx \sqrt{2}>1$.
\begin{figure}[ht]
\includegraphics[width=\columnwidth]{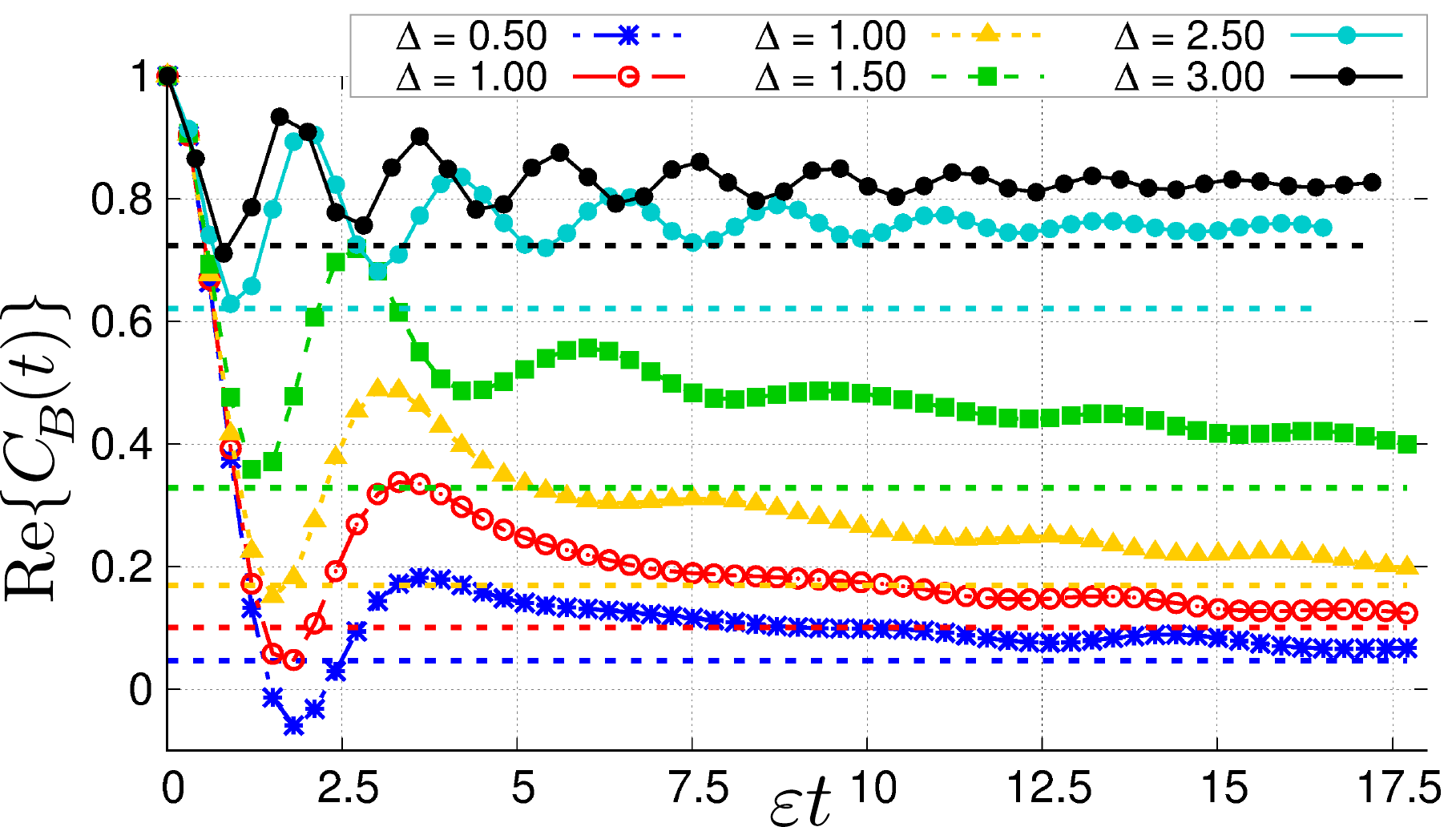} 
\caption{\label{fig:Ap4_} {\bf Fitting.} Non-renormalized SMCF for different values of $\Delta$. The dashed horizontal lines represent the values of $\langle\sigma^x\rangle^2$ corresponding to the renormalization factor to be substracted..}
\end{figure}
Notice that in the weak coupling solution Eq.~(\ref{weakfun}) there is a constant term $b_0$ that does not cancel with $\langle\sigma^x\rangle^2$. Indeed, the weak coupling initial state (\ref{separable}) is no longer in equilibrium, and gives rise to a time dependent $\langle\sigma^x\rangle\approx \ttr_E\{\rho_\sigma(t)\otimes\rho_R \sigma_x\}$ that not cancel out $b_0$.
In contrast, in the main text we analyze intermediate regimes of $r/\Delta$ and therefore we have to consider the full thermal initial state $\rho^\tth_E$, for which 
\bea
\langle\sigma^x\rangle= \ttr_E\{e^{iH_E t}\sigma^x e^{-iH_E t}\rho^\tth_E \}=\ttr_E\{ \sigma^x\rho^\tth_E \}, 
\eea
is time-independent. Moreover, Fig~\ref{fig:Ap4_} shows that the non-renormalized correlation function decays at long times to $\langle\sigma^x\rangle^2$ (dashed lines with the same color), at least for the time-scales reachable within our simulations. In other words, the non-renormalized correlation decays to a constant $b_0\approx \langle\sigma^x\rangle^2$ when $1/T_0\neq 0$, and to an offset value $\tilde C_0\rightarrow C_0$ when $1/T_0\rightarrow 0$.

\section{Appendix C: Correlation function in the harmonic limit}
Let us now consider the particular case of Gaussian environments. In this case, the environment should be formed by a set of independent harmonic oscillators, and thus we relabel the environment eigenstates and eigenenergies in terms of the standard indices: one referring to the oscillator, $\lambda$ and another one referring to its internal state $n_\lambda$ that reflects the number of quanta in such oscillator (we consider a single molecule and thus skip the index $j$),
\bea
&&|\epsilon_k\rangle\equiv |n_\lambda \rangle,\cr
&&\epsilon_k\equiv \epsilon_{n_\lambda}=\omega_\lambda(n_\lambda+1/2),
\eea
Thus, the correlation function can be written as
\bea
C_B(t)=\sum_{\lambda',n_{\lambda'}}\sum_{\lambda'',n_{\lambda''}}e^{-\beta\omega_{\lambda'}}|{B}^{n_{\lambda'},n_{\lambda''}}|^2e^{i(\epsilon_{n_{\lambda'}}-\epsilon_{n_{\lambda''}})t}.\nonumber
\eea
To obtain Gaussian statistics we need furthermore a linear coupling operator $B=\sum_\lambda g_\lambda (a_\lambda^\dagger+a_\lambda)$, such that ${B}^{n_{\lambda'},n_{\lambda''}}$ is 
\bea
\langle n_{\lambda'}|B|n_{\lambda''}\rangle&=&\sum_\lambda \delta_{\lambda',\lambda}\delta_{\lambda'',\lambda}g_\lambda\big(\sqrt{n_{\lambda''}+1}\delta_{n_{\lambda'},n_{\lambda''}+1}\cr
&+&\sqrt{n_{\lambda''}}\delta_{n_{\lambda'},n_{\lambda''}-1}\big).\nonumber
\eea
Notice that here there is no need to renormalize, since $\langle n_{\lambda}|B|n_\lambda\rangle=0$, i.e the coupling operator does not connect the same environment eigenstates, which means that $\tilde{B}=B$.
In addition, we have 
\bea
&&\delta_{n_{\lambda'},n_{\lambda''}+1}e^{i(\epsilon_{n_{\lambda'}}-\epsilon_{\lambda''})t}=\delta_{n_{\lambda'},n_{\lambda''}+1}e^{i\omega_\lambda t}\cr
&&\delta_{n_{\lambda'},n_{\lambda''}-1}e^{i(\epsilon_{\lambda'}-\epsilon_{\lambda''})t}=\delta_{n_{\lambda'},n_{\lambda''}-1}e^{-i\omega_\lambda t}.
\eea
Considering also the Bose-Einstein statistics, i.e. 
\bea
N(\omega_\lambda)=\frac{\sum^\infty_{n_\lambda=0}n_\lambda e^{-\beta \omega_\lambda(n_\lambda+1/2)}}{Z_\lambda}=\frac{1}{e^{-\beta\omega_\lambda}-1},
\eea
and the detailed balance $(N(\omega_\lambda)+1)e^{-\beta\omega_\lambda}=N(\omega_\lambda)$, we find the standard correlation function for harmonic environments,  
\bea
C(t)=\sum_\lambda g_\lambda^2[(N(\omega_\lambda)+1)e^{-i\omega_\lambda t}+N(\omega_\lambda) e^{i\omega_\lambda t}].
\eea
As it is well-known, this function can be written in terms of the one particle spectral density $J(\omega)=g(\omega)^2|\partial(\omega_k)/\partial k|^{-1}_{k=k(\omega)}$ as
\bea
C(t)=\int_0^\infty d\omega J(\omega)[(N(\omega)+1)e^{-i\omega t}+N(\omega)e^{i\omega t}].
\eea
As discussed in regard to Eq. (\ref{CE}), as long as $J(\omega)$ is a smooth function in frequencies (since $N(\omega)$ is smooth), $C(t)$ will decay to zero. This is the case in most physical models, where $J(\omega)$ is a continuous differentiable function. 

\section{Appendix D: Additional numerics on the offset}\label{App3}
\begin{figure}[ht]
\includegraphics[width=\columnwidth]{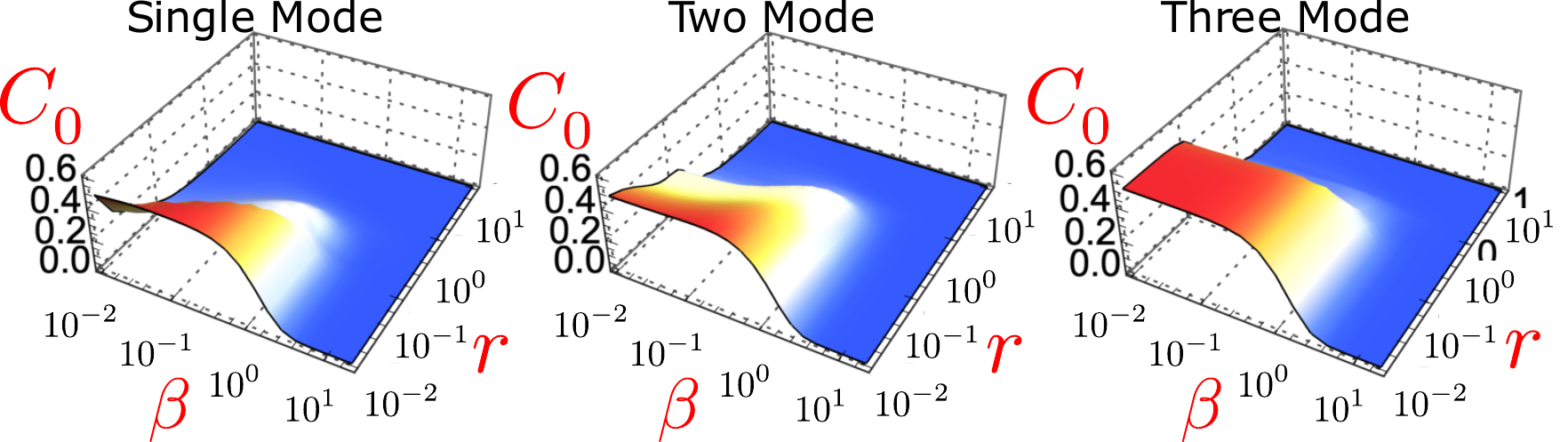} 
\caption{\label{fig:Ap1} {\bf Few-Mode Case: Offset.} Offset as a function of the inverse of the temperature $\beta$ and the environment interaction $g_{\lambda}\propto r$ for a single-, two- and three-mode molecule, with $\Delta=\varepsilon=1$.}
\end{figure}

We first compute the offset $C_0$ using Eq.~\ref{offset} by diagonalizing $H_{\rm SM}$ for a single, two and three model molecule. As seen in the main text, the offset does not show any strong dependence on the number of modes $L$, and it is a smooth function of the parameters of interest. The results is shown in~Fig.\ref{fig:Ap1} where it can be noticed that $x$- and $y$-axis are in log-scale. The figure also illustrates why the offset cancels towards the limit in which the environment Hamiltonian becomes separable, for instance, as the ratio $r/\Delta\gg 1$.
\begin{figure}[ht]
\includegraphics[width=\columnwidth]{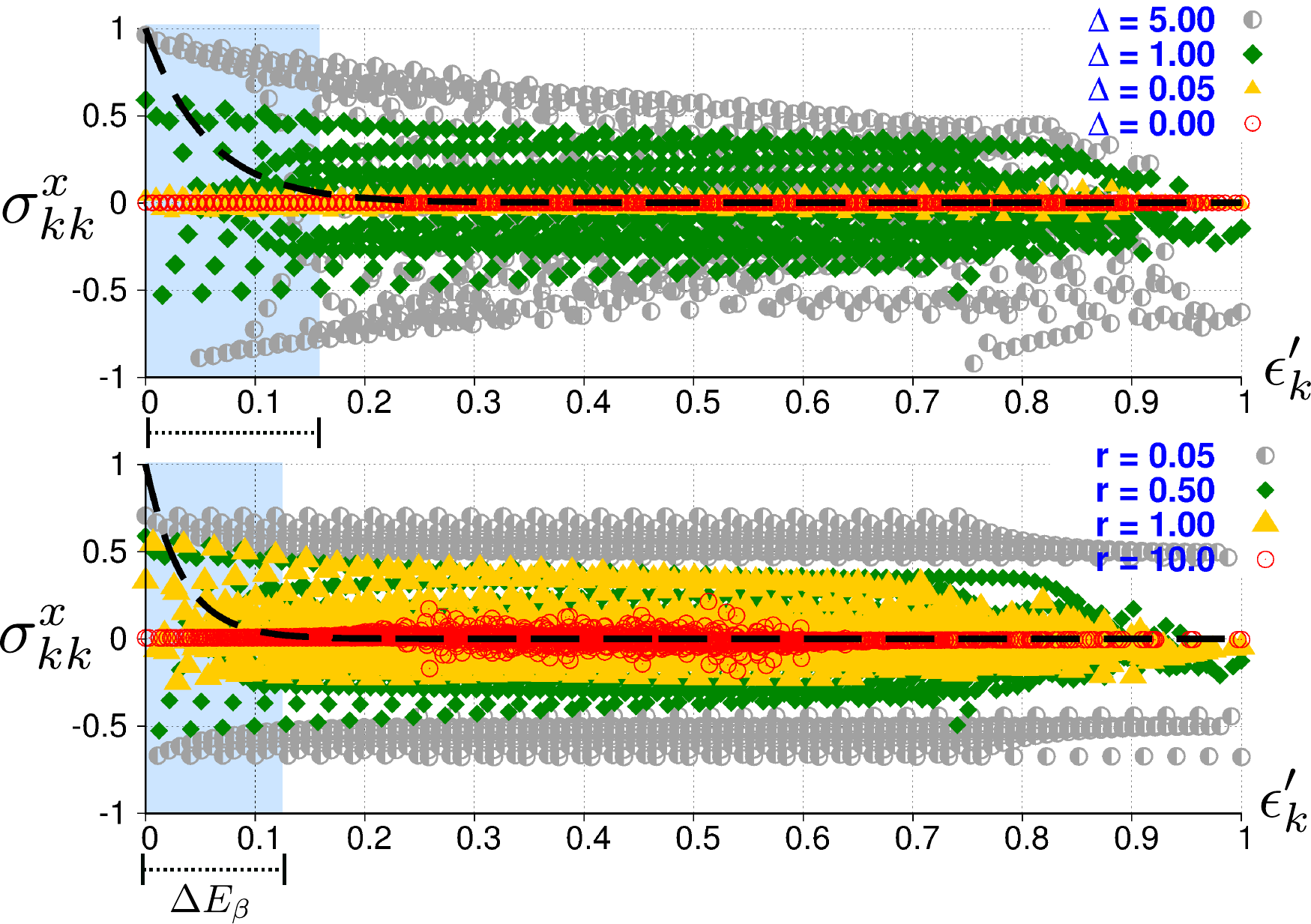} 
\caption{\label{fig:Ap2} {\bf Few-mode Case: Statistics.} Distribution of the single-eigenstate expectation of system-environment coupling operator $B=\sigma^x$ as a function of (a) the tunneling strength $\Delta$ with $r=0.25$, and (b) $r$ $(\propto g_{\lambda})$ with $\Delta=1$. The black dashed lines represent the thermal distribution for $\beta=0.5$.}
\end{figure}

In addition, we compute the distribution of $B^{kk}=\sigma_{kk}^x=\langle \epsilon_k|\sigma^x|\epsilon_k\rangle$. The eigenenergies are rescaled as $\epsilon'_k=(\epsilon_k-\epsilon_0)/{\rm max}(\{\epsilon_k\}-\epsilon_0)$, such that $0\leq \epsilon'_k\leq 1$. From this figure it is easily noticed that whenever $\Delta=0$, then $B^{kk}=0$ for all eigenstates, as expected, while for a very small value of $\Delta$ the distribution already gets spread around zero. The larger the value of $\Delta$, the broader the distribution, and as seen in Fig.~\ref{fig:Ap2}, when $\Delta/r\gg 1$ two peaks are getting formed around $\pm 1$, specially in the low-energy regime that is the relevant energy bandwidth for the computation of $C_0$. Such relevant bandwidth can be estimated by considering the thermal distribution $\eta_k$ (black dashed lines). 

Furthermore, to visualise the energy range in which the offset increases we compute an effective participation-energy range $\Delta E_{\beta}=\epsilon_N-\epsilon_0$ (light blue region), defined as the values in which the cumulative sum $F_\beta(N)=\sum^{N}_{k=0}\eta_k(B^{kk})^2$ converge into their constant value $F_\beta(N)\approx \lim_{N\rightarrow\infty}F_\beta(N)$. 
This quantity might be helpful to evaluate the offset for thermal large systems, where only the first few eigenstates are accessible. Nonetheless, as concluded before, a nonzero offset implies the participation of at least two eigenstates.
 
\begin{figure}[ht]
\includegraphics[width=\columnwidth]{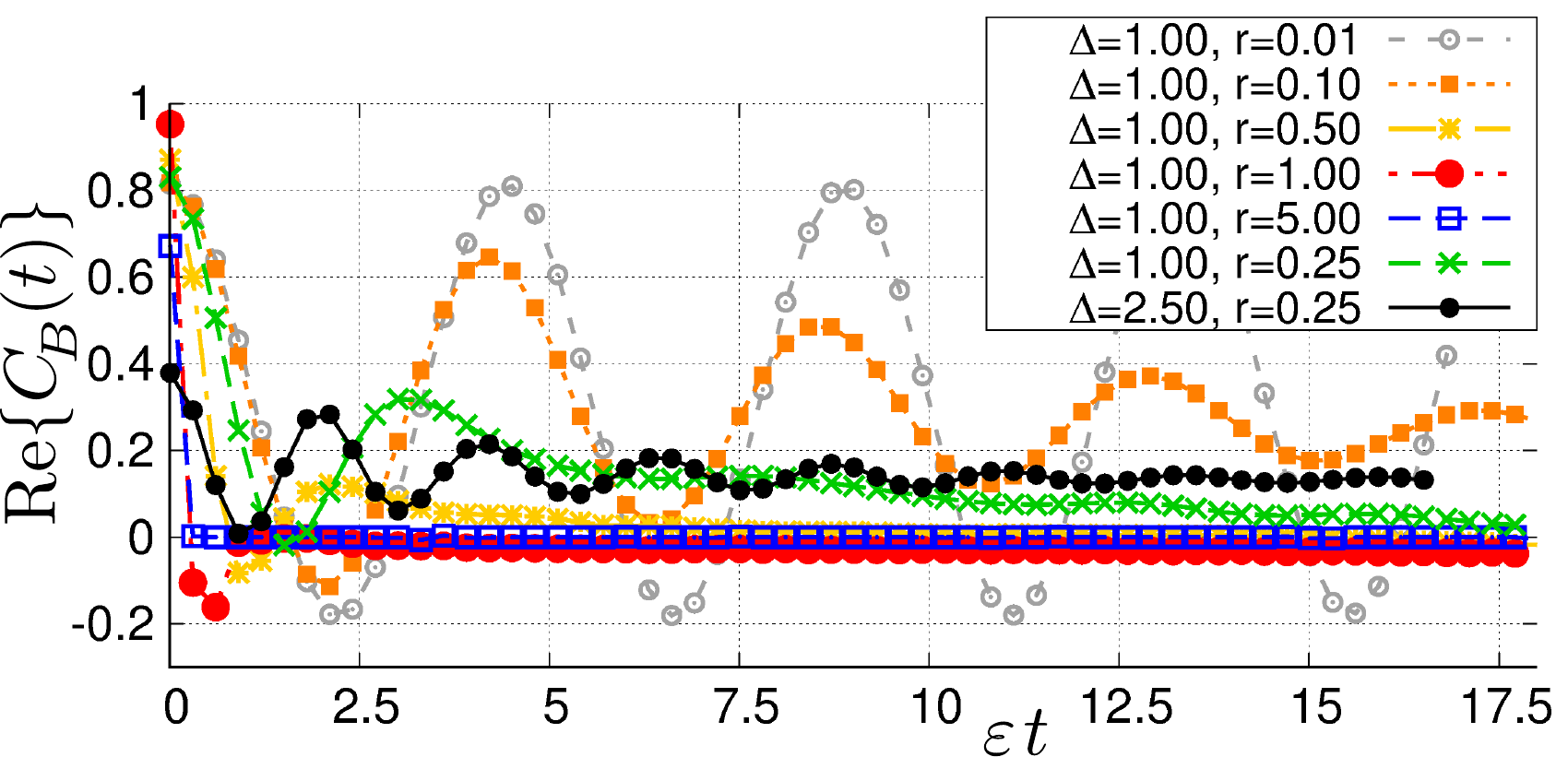} 
\caption{\label{fig:Ap3} {\bf Environment Interaction.} Single-molecule correlation function for different values of the environment interaction strength $r$, given $\Delta=1$ and $\beta=1$.}
\end{figure}

To complete the analysis, in Fig.~\ref{fig:Ap3} we show the variation of the SMCF as the interaction strength $r$ is increased. We can see that the larger the ratio $r/\Delta$  the more relevant the dephasing becomes and the faster the correlation function decays. To contrast these results, we additionally include the correlation function with parameters $r=0.25$ and $\Delta=2.5$ for which the appearance of a very large $T_0$ is observed.

\section{Appendix E: Many-Molecule Scenario}\label{App3}

A standard case is that in which the environment is composed of an ensemble of molecules. When they are independent, the total correlation function is given by $C_B(t)=\sum_{j=1}^M C_B^{(j)}(t)$. In a realistic setup the molecules in the ensemble are not identical, and the total correlation function shall depend on how the molecular parameters, $\{\Delta_j,\varepsilon_j\}$, are distributed. Let us assume that they follow a normalized distribution is ${\cal P}(\{\Delta_j,\varepsilon_j\})$. In the continuous limit, the correlation function can be then written as
\begin{eqnarray}
  C_B(t) &=& \frac{1}{M}\sum_j C_B(t,\{\Delta_j,\varepsilon_j\})\nonumber\\
  &=&\int_{<\Delta>}\int_{<\varepsilon>} {\cal P}(\Delta,\varepsilon)C_B(t,\Delta,\varepsilon)\,d\Delta\, d\varepsilon
\end{eqnarray}
with $C_B^{(j)}(t)\equiv C_B(t,\{\Delta_j,\varepsilon_j\})$, and ${\cal P}(\Delta,\varepsilon)$ is the continuous  distribution. A precise knowledge of this distribution may not be experimentally trivial to obtain. 

Here, we consider a Gaussian probability distribution where $\Delta$ and $\epsilon$ are treated as independent variables,
\begin{eqnarray}
  {\cal P}(\Delta,\varepsilon)&=&{\cal P}(\Delta){\cal P}(\varepsilon)\nonumber\\
  &=&\left(\frac{1}{\sqrt{2\pi \sigma^2}}\right)^2 e^{\frac{-(\Delta-\overline{\Delta_j})^2}{2\sigma^2}}e^{\frac{-(\varepsilon-\overline{\varepsilon_j})^2}{2\sigma^2}},
\end{eqnarray}
where the distribution width $\sigma$ is set as a tunning parameter. Considering such a Gaussian distribution (which leads to mean values $\{\overline{\Delta_j}=1,\overline{\varepsilon_j}=1\}$), we randomly sample the parameter set $\{\Delta_j,\varepsilon_j\}$ and compute the correlation function for each pair by solving the eigenspectrum of $H_E^{(j)}$.
\begin{figure}[ht]
\includegraphics[width=\columnwidth]{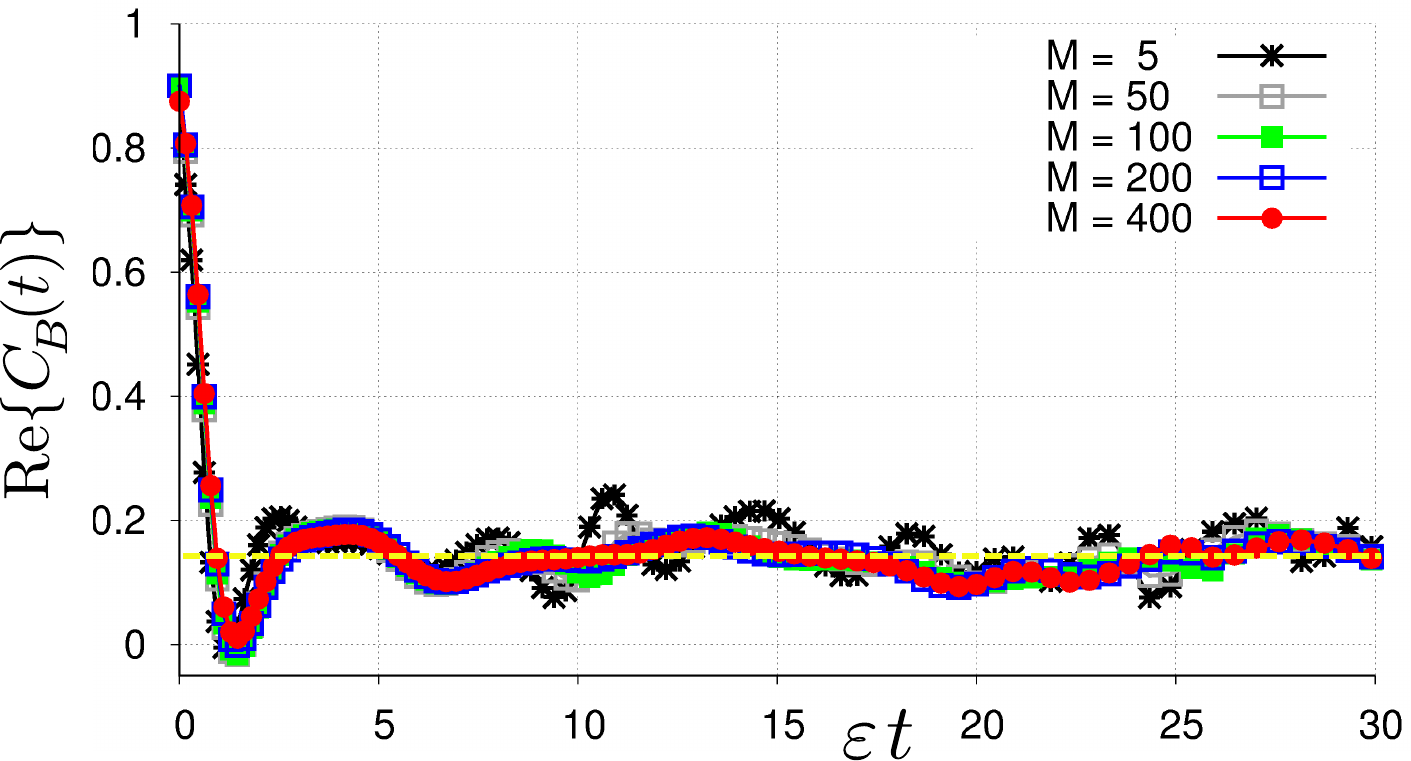} 
\caption{\label{fig:Ap4} {\bf Many-molecule Offset.} Many-molecules correlation function computed using the eigenstates and eigenvalues of $H_E^{(j)}$. The parameter are $r=0.25$, $\beta=1$, and we randomly choose $\{\Delta_j,\varepsilon_j\}$ around $\overline{\Delta_j}=\overline{\varepsilon_j}=1$, and a distribution width $\sigma = 0.3$.}
\end{figure}
 
In Fig.~\ref{fig:Ap4} we present the numerical results of the correlation function as a function of the total number of molecules $M$, each of them with a single \textit{active} bosonic mode. Therein, it can be seen that the function exhibits a large offset in presence of many molecules, and interestingly, the expected finite-size recurrence is washed out in the average. The fluctuations around the final offset value (light yellow dashed line) as expected to disappear as the number of molecules involved is larger while $\sigma$ also is increased. This calculation, for higher number of modes is not easily accessible due to the exponential growth of the Hilbert space of $H_E^{(j)}$. 

\subsection{The 1/f noise}
Another important aspect of the presence of an offset for some parameters is related to the band-width of the 1/f noise. To show this we compute the susceptibilty function, which is the Fourier transform of $C_B(t)$. 
The 1/f is observed at low frequencies, and therefore it will be dominated by the long time or slowly decaying term of the fitting correlation function (\ref{fit}), i.e. 
\bea
f^{\textmd{lt}}(t)=\tilde C_0 e^{-t/T_0}.
\label{slow}
\eea
In order to compute the susceptibility, one should take into account a sum over an ensemble of molecules $j$, each of them with different parameters and therefore with different decay rates $\nu_0^{(j)}=1/T_0^{(j)}$,
\bea
F^{\textmd{lt}}(t)=\sum_j \tilde C^{(j)}_0 e^{-\nu_0^{(j)}t}. 
\eea
Thus, the susceptibility at low frequencies will be given by the Fourier transform of this function, 
\begin{eqnarray}
  \chi_0(\omega) &=& \sum_j\int_{-\infty}^{\infty}e^{-\nu_0^{(j)}|t|}e^{-i\omega t} \tilde C_0^{(j)}dt\nonumber\\
  &=&\frac{c_1}{2\pi}\sum_j\tilde{C}_0^{(j)}\frac{\nu_0^{(j)}}{\omega^2+\frac{(\nu_0^{(j)})^2}{4\pi^2}}\nonumber\\
  &\propto& \frac{c_1}{2\pi}\int_{\nu_{\rm min}}^{\nu_{\rm max}}{\cal Q}(\nu_0) \tilde{C}_0(\nu_0)\frac{\nu_0}{\omega^2+\frac{\nu_0^2}{4\pi^2}}d\nu_0
\end{eqnarray}
with $\nu_{\rm min}=1/T_{\rm max}$ and $\nu_{\rm max}=1/T_{\rm min}$ very small. In the last equality, we have considered the limit of a very dense molecular ensemble and taken the sum as an integral with a certain probability distribution for decay rates, which we take as ${\cal Q}(\nu_0)=1/\nu$ \cite{milotti2002,faoro2004}. Moreover, we assume a narrow band-width at low frequencies, such that 
$\tilde{C}_0(\nu_0)\approx \tilde{C}_0$ so that it can be taken out of the integral. The resulting integral is analytically accessible and allows to obtain the 1/f behaviour, since
\begin{eqnarray}
\chi_0(\omega)\propto \frac{\tilde C_0}{\omega}\,,\;\;\;\text{if}\;\;\;\frac{1}{T_{\rm max}}<\nu_0<\frac{1}{T_{\rm min}}\,.
\end{eqnarray}
So far, we have carried the standard calculation of the 1/f spectrum. The novelty in our case is that due to the finite temperature, we find that $\nu_{\rm min}=1/T_{\rm max}=0$, since there are some molecules with internal parameters that have an infinitely decaying correlation function. Previous models are often based on a weak coupling approximation that lead to a correlation function (\ref{weakfun}) that always decays with a finite rate $\lambda$. This leads to a 1/f spectrum limited to the lower frequency $\frac{1}{T_{\rm max}}=\lambda_{\rm min}$. In contrast, our model shows a 1/f behaviour that extends to zero-frequency, and it is therefore in better agreement with experimental observations (see \cite{milotti2002} and references therein).
\end{document}